\documentclass[twocolumn,american,english]{revtex4}
\usepackage[T1]{fontenc}
\usepackage[latin9]{inputenc}
\usepackage{color}
\usepackage{array}
\usepackage{float}
\usepackage{multirow}
\usepackage{amsmath}
\usepackage{graphicx}

\makeatletter

\providecommand{\tabularnewline}{\\}

\@ifundefined{textcolor}{}
{%
 \definecolor{BLACK}{gray}{0}
 \definecolor{WHITE}{gray}{1}
 \definecolor{RED}{rgb}{1,0,0}
 \definecolor{GREEN}{rgb}{0,1,0}
 \definecolor{BLUE}{rgb}{0,0,1}
 \definecolor{CYAN}{cmyk}{1,0,0,0}
 \definecolor{MAGENTA}{cmyk}{0,1,0,0}
 \definecolor{YELLOW}{cmyk}{0,0,1,0}
}

\makeatother

\usepackage{babel}
\begin{document}

\title{\textcolor{black}{The dynamic electron-correlation energy in the
NOF-MP2 method from the orbital-invariant perturbation theory}}

\author{Mario Piris }

\address{Kimika Fakultatea, Euskal Herriko Unibertsitatea (UPV/EHU), P.K.
1072, 20080 Donostia, Euskadi (Spain);}

\address{Donostia International Physics Center (DIPC), 20018 Donostia, Euskadi
(Spain)}

\address{IKERBASQUE, Basque Foundation for Science, 48013 Bilbao, Euskadi
(Spain).\vspace{0.5cm}
}
\begin{abstract}
\textcolor{black}{The original formulation (Phys. Rev. Lett. 119,
063002, 2017) of the natural orbital functional - second-order Møller\textendash Plesset
(NOF-MP2) method is based on the MP2 that uses the canonical Hartree-Fock
molecular orbitals. The current work presents a reformulation of the
dynamic energy correction based on the orbital-invariant MP2, which
allows to attain both dynamic and static correlations even for those
systems with strong orbital localizability and significant multiconfigurational
character. To improve the reference Slater determinant formed with
natural orbitals, the natural orbital functional that generates them
is also modified to take into account only the inter-pair static correction.
This more general NOF-MP2 is able to dissociate properly noble gas
dimers, which remain as non-bound species within the canonical formulation.
Test calculations in a selected set of 30 polyatomic molecules demonstrate
a substantial improvement not only of the relative energies but also
of the total energies calculated with the NOF-MP2 method.}\vspace{0.3cm}

DOI: 10.1103/PhysRevA.98.022504

\end{abstract}

\maketitle
A reliable electronic structure method must be able to describe in
a balanced way both static (non-dynamic) and dynamic electron correlation\textcolor{black}{{}
\citep{Valderrama1997,Valderrama1999}}. Recently \citep{Piris2017},
a single-reference global method for electron correlation was introduced
taking as reference the Slater determinant of natural orbitals (NOs)
obtained from an approximate natural orbital functional (NOF) \citep{Piris2018}.
In this approach, the total energy is formed as $\tilde{E}_{hf}+E^{dyn}+E^{sta}$,
where $\tilde{E}_{hf}$ is the Hartree-Fock (HF) energy obtained with
NOs, the dynamic energy ($E^{dyn}$) is derived from a modified second-order
Møller\textendash Plesset perturbation theory (MP2), while the non-dynamic
energy ($E^{sta}$) is obtained from the static component of the NOF.

The success of the method, called NOF-MP2, is determined by the NOs
used to generate the reference. In \citep{Piris2017}, orbitals were
obtained from the Piris natural orbital functional 7 (PNOF7) there
proposed, an interacting-pair model that recovers the intra-pair but
only static inter-pair correlation. As a consequence, PNOF7 NOs can
be localized in certain regions of space, depending on the degree
of interaction between the electron pairs. When the inter-pair non-dynamic
correlation is negligible, these orbitals turn out to be close to
the known NOs of the independent-pair model (PNOF5) \citep{Piris2013}.
In general, NOs will be located in those regions where their atomic
orbitals responsible for the intra-pair and static correlation are
found. It is worth noting that localized NOs provide an orbital picture
with a clear chemical meaning \citep{Matxain2012a} that is not easy
to obtain using canonical orbitals.

On the other hand, $E^{dyn}$ was formulated \citep{Piris2017} from
traditional MP2 energy that involves the use of canonical HF molecular
orbitals, therefore, a reformulation of dynamic energy correction
is necessary so that any type of orbital can be used. Perturbation
theory with non-canonical orbitals (in most cases localized orbitals)
has been used formerly \citep{Saebo2002,Werner2015} in order to speed
up processing times. As Pulay pointed out \citep{Pulay1983}, the
increase in computational cost associated with the increase in the
number of electrons is not justified and is mainly due to the use
of canonical orbitals. In the last three decades, the orbital localizability
has been exploited by several approaches known as linear-scaling methods
\citep{LSTCQ2011}. The latter have extended the applicability of
wavefunction-based correlation methods to larger electronic systems.
Consequently, an additional motivation for a reformulation of $E^{dyn}$
is the possibility of computer savings.

\textcolor{black}{The present work pursues two objectives. On the
one hand, improve NOs with which the reference determinant is built
and, on the other hand, propose a correction $E^{dyn}$ based on the
orbital-invariant (oi) MP2 energy.  As a result, a new variant of
the method we will call NOF-oiMP2 emerges, whereas from now on we
will refer to the original version as NOF - canonical MP2 (NOF-cMP2).}

In NOF theory \citep{Piris2007b,Piris2014a}, the ground-state electronic
energy ($E$) is given in terms of the NOs $\{\phi_{i}\}$ and their
occupation numbers (ONs) $\{n_{i}\}$. Unfortunately, the exact reconstruction
$E[\{n_{i},\phi_{i}\}]$ has been an unattainable goal so far, therefore
we are talking about orbitals that diagonalize the one-particle reduced
density matrix (1-RDM) corresponding to an approximate ground-state
energy, and it is more appropriate to talk about NOF instead of a
1-RDM functional due to the existing dependence on the reconstructed
two-particle RDM (2-RDM).

Restrictions on the ONs to the range $0\leq n_{i}\leq1$ represent
the necessary and sufficient conditions for ensemble N-representability
of 1-RDM \citep{Coleman1963} under the normalization condition $\sum_{i}n_{i}=\mathrm{N}$.
Note that we focus on the N-representability problem for statistical
one-matrix ensembles, since to guarantee the pure-state N-representability
conditions \citep{Klyachko2006,Altunbulak2008} only 1-RDM ensemble
constraints are necessary if $E[\{n_{i},\phi_{i}\}]$ is a pure N-representable
functional \citep{Valone1980,Nguyen-Dang1985}.

In approximate one-particle theories, the 2-RDM plays a dominant role
that determines the functional N-representability \citep{Piris2018}.
The use of 2-RDM ensemble N-representability conditions \citep{Mazziotti2012}
for generating a reconstruction functional was proposed in Ref. \citep{Piris2006},
where auxiliary matrices $\mathbf{\triangle}$ and $\Pi$ were introduced
to reconstruct the two-particle cumulant \citep{Mazziotti1998}. In
this communication, we address only singlet states and adopt a restricted
spin theory, so that energy reads
\begin{equation}
\begin{array}{c}
E=2\sum\limits _{p}n_{p}\mathcal{H}_{pp}+\sum\limits _{qp}\Pi_{qp}\mathcal{L}_{pq}\qquad\qquad\\
+\sum\limits _{qp}\left(n_{q}n_{p}-\Delta_{qp}\right)\left(2\mathcal{J}_{pq}-\mathcal{K}_{pq}\right)
\end{array}\label{PNOF}
\end{equation}
where $\mathcal{H}_{pp}$ denotes the diagonal elements of the core-Hamiltonian,
while $\mathcal{J}_{pq}$, $\mathcal{K}_{pq}$, and $\mathcal{L}_{pq}$
are the direct, exchange, and exchange-time-inversion integrals \citep{Piris1999}.
Appropriate forms of matrices $\Delta$ and $\Pi$ lead to different
implementations known in the literature as PNOFi (i=1-7) \citep{Piris2013b,Piris2014c,Piris2017}.
Remarkable is the case of PNOF5 \citep{Piris2011,Piris2013e} which
turned out to be pure N-representable \citep{Pernal2013,Piris2013c}.

The conservation of the total spin allows to derive the diagonal elements
$\Delta_{pp}=n_{p}^{2}$ and $\Pi_{pp}=n_{p}$ \citep{Piris2009}.
The 2-RDM N-representability $D$ and $Q$ conditions lead to inequalities
$\Delta_{qp}\leq n_{q}n_{p}$ and $\Delta_{qp}\leq h_{q}h_{p}$ \citep{Piris2006},
where $h_{p}=1-n_{p}$. To fulfill the $G$ condition, the off-diagonal
elements of the $\Pi$-matrix must satisfy the constraint \citep{Piris2010a}
\begin{equation}
\Pi_{qp}^{2}\leq\left(n_{q}h_{p}+\Delta_{qp}\right)\left(h_{q}n_{p}+\Delta_{qp}\right)\label{G_cond}
\end{equation}
For a given approximation of $\Delta_{qp}$, it is evident that the
modulus of $\Pi_{qp}$ is determined from Eq. (\ref{G_cond}) assuming
the equality, however, there is no hint to determine the sign of $\Pi_{qp}$.
The requirement that for any two-electron singlet the NOF (\ref{PNOF})
yields the accurate energy expression obtained from the exact wavefunction
\citep{Lowdin1955d}, implies \citep{Piris2010a} that $\Delta_{qp}=n_{q}n_{p}$
and $\left|\Pi_{qp}\right|=\sqrt{n_{q}n_{p}}$, respectively. Furthermore,
the phase factor of $\Pi_{qp}$ can be $+1$ if $q,p\in\left(1,\infty\right)$,
and -1 otherwise. 

To achieve a model of independent pairs with N>2, the orbital space
$\Omega$ is divided into N/2 mutually disjoint subspaces $\Omega{}_{g}$,
so each subspace contains one orbital $g$ below the level N/2, and
$\mathrm{N}_{g}$ orbitals above it, which is reflected in additional
sum rules for the ONs ($\sum n_{p}=1,p\in\Omega_{g}$). In what follows,
let's consider $N_{g}$ equal to a fixed number that corresponds to
the maximum value allowed by the basis set used. Keeping $\Delta_{qp}=n_{q}n_{p}$,
and generalizing the two-electron expression for off-diagonal elements
of $\Pi$-matrix, namely, $\Pi_{qp}^{g}=\sqrt{n_{q}n_{p}}$ if $q,p>\mathrm{N}/2$,
and $\Pi_{qp}^{g}=-\sqrt{n_{q}n_{p}}$ if \textit{q}=\textit{g} or
\textit{p}=\textit{g}, we obtain the extended PNOF5 \citep{Piris2013e}. 

In Ref. \citep{Piris2017}, non-zero $\Pi_{qp}$ elements were considered
among orbitals belonging to different subspaces \citep{Piris2017},
whereas $\mathrm{\Delta_{\mathit{qp}}=0}$. From Eq. (\ref{G_cond})
follows that provided the $\Delta_{qp}$ vanishes, $\left|\Pi_{qp}\right|\leq\Phi_{q}\Phi_{p}$
with $\Phi_{q}=\sqrt{n_{q}h_{q}}$. Assuming equality, and generalization
of the sign convention adopted for extended PNOF5, i.e., $\Pi_{qp}^{\Phi}=\Phi_{q}\Phi_{p}$
if $q,p>\mathrm{N}/2$, and $\Pi_{qp}^{\Phi}=-\Phi_{q}\Phi_{p}$ otherwise,
led to PNOF7 \citep{Piris2017}.

Another possible option, that favors decreasing of the energy (\ref{PNOF}),
is to consider all the inter-pair factors negative, ergo, $\Pi_{qp}^{\Phi}=-\Phi_{q}\Phi_{p}$.
Recently \citep{mitxelena2018a}, we have analyzed several examples
with strong static correlation, specifically, the one-dimensional
Hubbard model with up to 14 sites and rings with up to 16 hydrogens.
Comparing with accurate diagonalization calculations, our results
indicate that all negative inter-pair factors is a better option.

In addition, it would be convenient to take into account the inter-pair
static correction in the NOF from the outset, thus preventing the
ONs and NOs from suffering an inter-pair non-dynamic influence, however
small, in the dynamic correlation domains. Taking into account the
$fg$-th inter-pair static correlation energy \citep{Piris2017},
\begin{equation}
E_{fg}^{sta}=\sum\limits _{p\in\Omega_{f}}\sum\limits _{q\in\Omega_{g}}4\Phi_{p}\Phi_{q}\,\Pi_{qp}^{\Phi}\mathcal{\,L}_{pq}=\sum\limits _{p\in\Omega_{f}}\sum\limits _{q\in\Omega_{g}}\Pi_{qp}^{s}\mathcal{\,L}_{pq}\,,\label{Esta_fg}
\end{equation}
we attain the new NOF:
\begin{equation}
\begin{array}{c}
E=\sum\limits _{g=1}^{\mathrm{N}/2}\sum\limits _{p\in\Omega_{g}}\left[n_{p}\left(2\mathcal{H}_{pp}+\mathcal{J}_{pp}\right)+\right.\sum\limits _{q\in\Omega_{g},q\neq p}\left.\Pi_{qp}^{g}\mathcal{L}_{pq}\right]\\
+\sum\limits _{f\neq g}^{\mathrm{N}/2}\sum\limits _{p\in\Omega_{f}}\sum\limits _{q\in\Omega_{g}}\left[n_{q}n_{p}\left(2\mathcal{J}_{pq}-\mathcal{K}_{pq}\right)+\Pi_{qp}^{s}\mathcal{L}_{pq}\right]
\end{array}\label{PNOF7s}
\end{equation}
where $\Pi_{qp}^{s}=-4n_{q}h_{q}n_{p}h_{p}$. This new approach will
henceforth refer to as PNOF7s and will provide the reference NOs to
form $\tilde{E}_{hf}$ in the NOF-oiMP2 method. The \textquotedbl s\textquotedbl{}
emphasizes that this interacting-pair model takes into account only
the static correlation between pairs, and therefore avoids double
counting in the regions where the dynamic correlation predominates,
already in the NOF optimization.

\begin{table*}
\centering{}\caption{\label{tab:ComparisonPNOF7}Comparison between PNOF7 and PNOF7s using
the cc-pVTZ basis set. $^{(a)}$ aug-cc-pVTZ was used.}
\vspace{0.3cm}
\begin{tabular}{ccccccccc}
\hline 
\multirow{2}{*}{Molecule} & \multirow{2}{*}{\qquad{}} & \multicolumn{3}{c}{$\mathrm{R_{\mathit{e}}}$ ($\textrm{\AA}$)} & \qquad{} & \multicolumn{3}{c}{$\mathrm{D_{\mathit{e}}}$ (kcal/mol)}\tabularnewline
 &  & PNOF7 & \enskip{}PNOF7s\enskip{} & Exp. &  & PNOF7 & \enskip{}PNOF7s\enskip{} & Exp.\tabularnewline
\hline 
H$_{2}$ &  & 0.743 & 0.743 & 0.743 &  & 108.6 & 108.6 & 109.5\tabularnewline
LiH &  & 1.604 & 1.603 & 1.595 &  & \enskip{}56.1 & \enskip{}56.4 & \enskip{}58.0\tabularnewline
Li$_{2}$ &  & 2.667 & 2.644 & 2.673 &  & \enskip{}23.3 & \enskip{}23.4 & \enskip{}24.4\tabularnewline
BH &  & 1.232 & 1.228 & 1.232 &  & \enskip{}75.7 & \enskip{}81.0 & \enskip{}81.5\tabularnewline
OH$^{-(a)}$ &  & 0.966 & 0.961 & 0.964 &  & \enskip{}87.0 & \enskip{}93.6 & -\tabularnewline
HF &  & 0.915 & 0.918 & 0.917 &  & 106.7 & 114.4 & 141.1\tabularnewline
LiF &  & 1.576 & 1.561 & 1.564  &  & \enskip{}95.4 & 104.6 & 139.0\tabularnewline
N$_{2}$ &  & 1.097 & 1.089 & 1.098  &  & 188.9 & 181.2 & 228.3\tabularnewline
CN$^{-(a)}$ &  & 1.186 &  1.169 & 1.177 &  & 212.0 & 202.7 & 240.7\tabularnewline
CO &  & 1.120 & 1.115 & 1.128  &  & 178.1 & 191.4 & 259.3\tabularnewline
NO$^{+}$ &  & 1.056 & 1.048 & 1.063 &  & 179.9 & 189.8 & -\tabularnewline
F$_{2}$ &  & 1.579 & 1.502 & 1.412  &  & \enskip{}\enskip{}2.6 & \enskip{}10.1 & \enskip{}39.2\tabularnewline
\hline 
\end{tabular}
\end{table*}

Like PNOF7, PNOF7s produces qualitatively correct potential energy
curves (PECs) for the dozen diatomic molecules studied in reference
\citep{Piris2017}. These systems cover a wide range of values for
binding energies ($\mathrm{D_{\mathit{e}}}$) and bond lengths ($\mathrm{R_{\mathit{e}}}$),
however, in all cases the correct dissociation limit implies an homolytic
cleavage of the bond with high degree of degeneracy effects. In Table
\ref{tab:ComparisonPNOF7}, a comparison between both functionals
is shown. The experimental bond lengths are taken from the National
Institute of Standards and Technology (NIST) Database \citep{nist},
whereas the experimental dissociation energies result from a combination
of Refs. \citep{nist} and \citep{Chase1998}. The correlation-consistent
valence triple-$\zeta$ basis set (cc-pVTZ) developed by Dunning \citep{Dunning1989}
was used throughout, except for the anionic species where the augmented
basis set (aug-cc-pVTZ) was used.

Table \ref{tab:ComparisonPNOF7} shows a slight shortening of the
equilibrium distances obtained with PNOF7s compared to those obtained
with PNOF7, whereas the dissociation energies experience a slight
increase, except in the cases of N$_{2}$ and CN$^{-}$. These minor
effects are related to the prevention of considering non-dynamic correlation
between pairs in the equilibrium regions where the dynamic correlation
prevails, and should lead to an improvement in the $\mathrm{R_{\mathit{e}}}$
and $\mathrm{D_{\mathit{e}}}$ calculated with the NOF-MP2 method.
As was pointed out in \citep{Piris2017}, the results are in good
agreement with the experiment for the smaller diatomics, for which
the electron correlation effect is almost entirely intrapair. When
the number of pairs increases, the theoretical values deteriorate
especially for the dissociation energies. This is related to a better
description of the asymptotic region with respect to the equilibrium,
therefore it is necessary to add the dynamic electron correlation
between pairs.

Now we focus on the reformulation of $E^{dyn}$. In the mid eighties,
Pulay and Saebø introduced an orbital invariant formulation of MP2,
the details of which can be found elsewhere \citep{Pulay1986,Saebo2002}.
The first-order wavefunction is a linear combination of all doubly
excited configurations, and their amplitudes $T_{pq}^{fg}$ are obtained
by solving the equations for the MP2 residuals. The MP2 energy correction
takes the form
\begin{equation}
E^{\left(2\right)}=\sum\limits _{g,f=1}^{\mathrm{N}/2}\sum\limits _{p,q>N/2}^{M}\left\langle gf\right|\left.pq\right\rangle \left[2T_{pq}^{gf}\right.\left.-T_{pq}^{fg}\right]\label{E2}
\end{equation}

where $M$ is the number of basis functions, and $\left\langle gf\right|\left.pq\right\rangle $
are the matrix elements of the two-particle interaction.

In NOF-cMP2, $E^{dyn}$ is obtained as the canonical $E^{\left(2\right)}$
modified to avoid double counting of the electron correlation \citep{Piris2017}.
The latter is divided into intra- and inter-pair contributions, and
the amount of dynamic correlation in each orbital $p$ is defined
by functions $C_{p}$ of its occupancy, namely,
\begin{equation}
\begin{array}{c}
C_{p}^{intra}=\begin{cases}
\begin{array}{c}
\begin{array}{c}
1-4h_{p}^{2}\end{array}\\
1-4n_{p}^{2}
\end{array} & \begin{array}{c}
p\leq\mathrm{N}/2\\
p>\mathrm{N}/2
\end{array}\end{cases}\\
\:C_{p}^{inter}=\begin{cases}
\begin{array}{c}
\begin{array}{c}
1\end{array}\\
1-4h_{p}n_{p}
\end{array} & \begin{array}{c}
p\leq\mathrm{N}/2\\
p>\mathrm{N}/2
\end{array}\end{cases}
\end{array}\label{Cp}
\end{equation}
According to Eq.(\ref{Cp}), fully occupied and empty orbitals yield
a maximal contribution to dynamic correlation, whereas orbitals with
half occupancies contribute nothing. It is worth noting that $C_{p}^{inter}$
is not considered if the orbital is below $\mathrm{N}/2$. Using these
functions as the case may be (intra-pair or inter-pair), we define
modified off-diagonal elements of the Fock matrix ($\tilde{\mathcal{F}}$)
as
\begin{equation}
\tilde{\mathcal{F}}_{pq}=\begin{cases}
C_{p}^{intra}C_{q}^{intra}\mathcal{F}_{pq}, & p,q\in\Omega_{g}\\
C_{p}^{inter}C_{q}^{inter}\mathcal{F}_{pq}, & otherwise
\end{cases}
\end{equation}
as well as modified two-electron integrals:
\begin{equation}
\widetilde{\left\langle pq\right|\left.rt\right\rangle }=\begin{cases}
C_{p}^{intra}C_{q}^{intra}C_{r}^{intra}C_{t}^{intra}\left\langle pq\right|\left.rt\right\rangle , & p,q,r,t\in\Omega_{g}\\
C_{p}^{inter}C_{q}^{inter}C_{r}^{inter}C_{t}^{inter}\left\langle pq\right|\left.rt\right\rangle , & otherwise
\end{cases}
\end{equation}
where the subspace index $g=1,...,N/2$. This leads to the following
linear equation for the modified MP2 residuals
\begin{equation}
\begin{array}{c}
\tilde{R}_{ab}^{ij}=\widetilde{\left\langle ab\right|\left.ij\right\rangle }+\left(\mathcal{F}_{aa}\right.+\mathcal{F}_{bb}-\mathcal{F}_{ii}-\left.\mathcal{F}_{jj}\right)T_{ab}^{ij}\:+\qquad\qquad\\
\\
{\displaystyle \sum_{c\neq a}\mathcal{\tilde{F}}_{ac}T_{cb}^{ij}}+{\displaystyle \sum_{c\neq b}}T_{ac}^{ij}\mathcal{\tilde{F}}_{cb}-{\displaystyle \sum_{k\neq i}}\tilde{\mathcal{F}}_{ik}T_{ab}^{kj}-{\displaystyle \sum_{k\neq j}}T_{ab}^{ik}\mathcal{\tilde{F}}_{kj}=0
\end{array}\label{residual}
\end{equation}
where $i,j,k$ refer to the strong occupied NOs, and $a,b,c$ to weak
occupied ones. It should be noted that diagonal elements of the Fock
matrix ($\mathcal{F}$) are not modified. 

By solving the linear system of equations (\ref{residual}) the amplitudes
$T_{pq}^{fg}$ are obtained, which are inserted into the Eq. (\ref{E2})
to achieve $E^{dyn}=E^{\left(2\right)}$. Following Ref. \citep{Piris2017},
the total energy of the system will be given by
\begin{equation}
E=\tilde{E}_{hf}+E^{corr}=\tilde{E}_{hf}+E^{sta}+E^{dyn}\label{Etotal}
\end{equation}
where $\tilde{E}_{hf}$ is the HF energy obtained with the NOs of
PNOF7s, Eq. (\ref{PNOF7s}), and $E^{sta}$ is the sum of energies
(\ref{Esta_fg}), 
\begin{equation}
E_{inter}^{sta}=\sum\limits _{f\neq g}^{\mathrm{N}/2}E_{fg}^{sta}=\sum\limits _{f\neq g}^{\mathrm{N}/2}\sum\limits _{p\in\Omega_{f}}\sum\limits _{q\in\Omega_{g}}\Pi_{qp}^{s}\mathcal{\,L}_{pq}\,,\label{Esta_inter}
\end{equation}
plus the static intra-pair electron correlation energy \citep{Piris2017},
\begin{equation}
E_{intra}^{sta}=\sum\limits _{g=1}^{\mathrm{N}/2}\sum\limits _{q\neq p}\sqrt{\Lambda_{q}\Lambda_{p}}\,\Pi_{qp}^{g}\mathcal{\,L}_{pq}\label{Esta_intra}
\end{equation}
In Eq. (\ref{Esta_intra}), note that $q,p\in\Omega_{g}$, and $\Lambda_{p}=1-\left|1-2n_{p}\right|$
is the amount of intra-pair static electron correlation in each orbital
$p$ as a function of its occupancy.

\textcolor{black}{The performance of NOF-oiMP2 has been tested in
several examples. Let's start with noble-gas dimers, which are held
together by dispersion, a manifestation of long-range dynamic correlation.
These species are not bound at the PNOF7 level of theory, and they
remain so even after adding $E^{dyn}$ using the canonical formulation.
With the new formulation of $E^{dyn}$ based on the orbital-invariant
MP2, the orbital localizability in noble-gas atoms can now be taking
into account, so that NOF-oiMP2 predicts bound species.}

\begin{figure}[H]
\caption{\label{fig:PEC-NG}Potential energy curves \textcolor{black}{of noble-gas
dimers} calculated at the NOF-oiMP2/aug-cc-pVTZ level of theory. The
zero-energy point has been set at 10 Å for each system.}
\vspace{0.8cm}

\centering{}\includegraphics[scale=0.3]{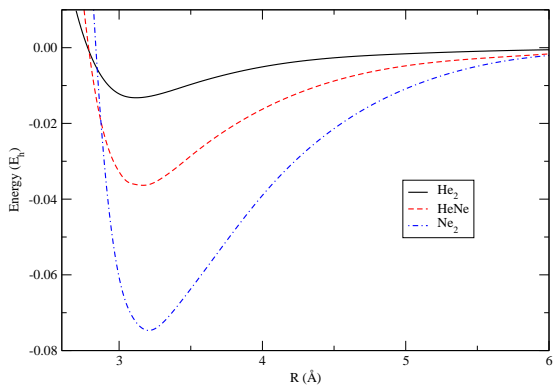}
\end{figure}

\begin{table}[h]
\caption{\label{tab:NobleGasDimers}Comparison of R$_{e}$($\textrm{\AA}$)
and D$_{e}$(kcal/mol) calculated at the MP2 and NOF-oiMP2 levels
of theory with the experimental values. The aug-cc-pVTZ basis set
was used.}
\vspace{0.3cm}
\centering{}%
\begin{tabular}{l|cc|cc|cc}
\hline 
\multirow{2}{*}{Dimer} & \multicolumn{2}{c|}{MP2} & \multicolumn{2}{c|}{NOF-oiMP2} & \multicolumn{2}{c}{Experiment}\tabularnewline
\cline{2-7} 
 & $\quad\mathrm{R_{\mathit{e}}\quad}$ & $\mathrm{\quad D_{\mathit{e}}\quad}$ & $\mathrm{\quad R_{\mathit{e}}\quad}$ & $\mathrm{\quad D_{\mathit{e}}\quad}$ & $\mathrm{\quad R_{\mathit{e}}}\quad$ & $\mathrm{\quad D_{\mathit{e}}\quad}$\tabularnewline
\hline 
He$_{2}$ & 3.09 & 0.013 & 3.12 & 0.013 & 2.97 & 0.022\tabularnewline
HeNe & 3.12 & 0.038 & 3.17 & 0.035 & 3.03 & 0.041\tabularnewline
Ne$_{2}$ & 3.18 & 0.076 & 3.21 & 0.074 & 3.09 & 0.084\tabularnewline
\hline 
\end{tabular}
\end{table}

\textcolor{black}{The potential energy curves (PECs) of He$_{2}$,
HeNe and Ne$_{2}$ are depicted in Fig.\ref{fig:PEC-NG}. For each
of the curves, the zero-energy point has been set at their corresponding
energy at 10 Å. It can be seen that NOF-oiMP2 produces qualitatively
correct PECs. In Table \ref{tab:NobleGasDimers}, the equilibrium
bond lengths (R$_{e}$) and dissociation energies (D$_{e}$) at the
MP2 and NOF-oiMP2 levels of theory can be found. The experimental
values were taken from Ref. \citep{Ogilvie1992}. The augmented correlation-consistent
valence triple-$\zeta$ basis set (aug-cc-pVTZ) \citep{Kendall1992,Woon1994}
was used in theoretical calculations. It is worth noting that larger
basis set is needed to adequately compare them with the experiment.
In addition, only valence electrons have been included in the correlation
treatment. It can be observed that both methods underestimate the
binding energies and overestimate the equilibrium distances, being
these effects more perceptible for the NOF-oiMP2. He2 is the worst
case since only a 60\% of the binding energy is recovered, while for
the other two systems it is between 85-92\%.}

\begin{figure}[H]
\caption{\label{fig:PEC-De}Potential energy curves with homolytic cleavage
of the bond calculated at the NOF-oiMP2/cc-pVTZ level of theory. The
zero-energy point has been set at 10 Å for each system.}
\vspace{0.8cm}
\centering{}\includegraphics[scale=0.3]{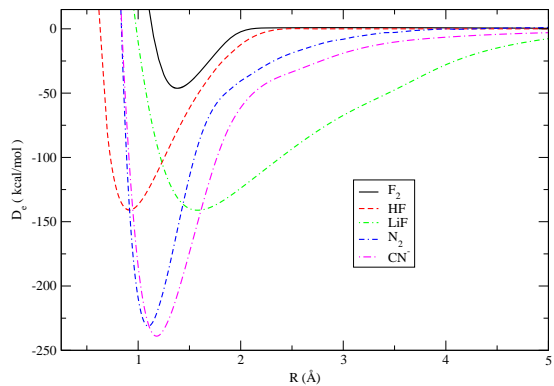}
\end{figure}

\begin{table*}
\caption{\label{tab:Comparison-De}Comparison between NOF-cMP2 and NOF-oiMP2
using the cc-pVTZ basis set. $^{(a)}$aug-cc-pVTZ was used.}
\vspace{0.3cm}
\centering{}%
\begin{tabular}{ccccccccc}
\hline 
\multirow{2}{*}{Molecule} & \multirow{2}{*}{\qquad{}} & \multicolumn{3}{c}{$\mathrm{R_{\mathit{e}}}$ ($\textrm{\AA}$)} & \qquad{} & \multicolumn{3}{c}{$\mathrm{D_{\mathit{e}}}$ (kcal/mol)}\tabularnewline
 &  & NOF-cMP2 & NOF-oiMP2 & Exp. &  & NOF-cMP2 & NOF-oiMP2 & Exp.\tabularnewline
\hline 
F$_{2}$ &  & 1.397 & 1.382 & 1.412  &  & \enskip{}34.5 & \enskip{}46.0 & \enskip{}39.2\tabularnewline
HF &  & 0.924 & 0.916 & 0.917 &  & 139.4 & 140.9 & 141.1\tabularnewline
LiF &  & 1.614 & 1.579 & 1.564  &  & 140.7 & 141.1 & 139.0\tabularnewline
N$_{2}$ &  & 1.084 & 1.098 & 1.098  &  & 224.2 & 230.7 & 228.3\tabularnewline
CN$^{-(a)}$ &  & 1.180 & 1.180 & 1.177 &  & 238.6 & 239.0 & 240.7\tabularnewline
\hline 
\end{tabular}
\end{table*}

There are no significant differences between the results obtained
with NOF-cMP2 and NOF-oiMP2 methods for diatomic systems analyzed
in Table \ref{tab:ComparisonPNOF7}. Representative PECs of these
molecules are depicted in Fig. \ref{fig:PEC-De}. Table \ref{tab:Comparison-De}
collects the electronic properties previously analyzed for systems
showed in Fig. \ref{fig:PEC-De}. The data reveals an outstanding
improvement in the dissociation energies with respect to PNOF7 and
PNOF7s, respectively. A slight improvement of the theoretical equilibrium
distances calculated with NOF-oiMP2 is also observed over those obtained
with NOF-cMP2.

\textcolor{black}{}
\begin{table*}
\begin{centering}
\textcolor{black}{\caption{\label{tab:Comparison}\textcolor{black}{Comparison of total electronic
energies, in Hartrees, calculated using the cc-pVTZ basis set at the
experimental geometry.}}
}
\par\end{centering}
\centering{}\textcolor{black}{}%
\vspace{0.3cm}
\begin{tabular}{ccccrrrrr}
\hline 
\textcolor{black}{No. } & \qquad{} & \textcolor{black}{Molecule } & \qquad{} & \textcolor{black}{NOF-cMP2} & \qquad{} & \textcolor{black}{NOF-oiMP2} & \qquad{} & \textcolor{black}{MP2\quad{}}\tabularnewline
\hline 
\textcolor{black}{\small{}1 } &  & \textcolor{black}{\small{}H$_{2}$O } &  & \textcolor{black}{\small{}-76.316438} &  & \textcolor{black}{\small{}-76.317906} &  & \textcolor{black}{\small{}-76.320480}\tabularnewline
\textcolor{black}{\small{}2 } &  & \textcolor{black}{\small{}NH$_{3}$ } &  & \textcolor{black}{\small{}-56.447022} &  & \textcolor{black}{\small{}-56.452165} &  & \textcolor{black}{\small{}-56.454549}\tabularnewline
\textcolor{black}{\small{}3 } &  & \textcolor{black}{\small{}CH$_{4}$ } &  & \textcolor{black}{\small{}-40.404095} &  & \textcolor{black}{\small{}-40.411188} &  & \textcolor{black}{\small{}-40.412721}\tabularnewline
\textcolor{black}{\small{}4 } &  & \textcolor{black}{\small{}HCN } &  & \textcolor{black}{\small{}-93.212795} &  & \textcolor{black}{\small{}-93.216612} &  & \textcolor{black}{\small{}-93.223664}\tabularnewline
\textcolor{black}{\small{}5 } &  & \textcolor{black}{\small{}C$_{2}$H$_{2}$ } &  & \textcolor{black}{\small{}-77.148255} &  & \textcolor{black}{\small{}-77.154876} &  & \textcolor{black}{\small{}-77.160778}\tabularnewline
\textcolor{black}{\small{}6 } &  & \textcolor{black}{\small{}PH$_{3}$ } &  & \textcolor{black}{\small{}-342.643167} &  & \textcolor{black}{\small{}-342.657773} &  & \textcolor{black}{\small{}-342.661029}\tabularnewline
\textcolor{black}{\small{}7 } &  & \textcolor{black}{\small{}Si$_{2}$H$_{6}$ } &  & \textcolor{black}{\small{}-581.624090} &  & \textcolor{black}{\small{}-581.641714} &  & \textcolor{black}{\small{}-581.643376}\tabularnewline
\textcolor{black}{\small{}8 } &  & \textcolor{black}{\small{}H$_{2}$CO } &  & \textcolor{black}{\small{}-114.288902} &  & \textcolor{black}{\small{}-114.301503} &  & \textcolor{black}{\small{}-114.309339}\tabularnewline
\textcolor{black}{\small{}9 } &  & \textcolor{black}{\small{}H$_{2}$S } &  & \textcolor{black}{\small{}-398.885411} &  & \textcolor{black}{\small{}-398.903604} &  & \textcolor{black}{\small{}-398.907289}\tabularnewline
\textcolor{black}{\small{}10 } &  & \textcolor{black}{\small{}C$_{2}$H$_{4}$ } &  & \textcolor{black}{\small{}-78.378826} &  & \textcolor{black}{\small{}-78.397165} &  & \textcolor{black}{\small{}-78.401267}\tabularnewline
\textcolor{black}{\small{}11 } &  & \textcolor{black}{\small{}CH$_{3}$OH } &  & \textcolor{black}{\small{}-115.494590} &  & \textcolor{black}{\small{}-115.514704} &  & \textcolor{black}{\small{}-115.519139}\tabularnewline
\textcolor{black}{\small{}12 } &  & \textcolor{black}{\small{}H$_{2}$O$_{2}$ } &  & \textcolor{black}{\small{}-151.308915} &  & \textcolor{black}{\small{}-151.325929} &  & \textcolor{black}{\small{}-151.334177}\tabularnewline
\textcolor{black}{\small{}13 } &  & \textcolor{black}{\small{}BF$_{3}$ } &  & \textcolor{black}{\small{}-324.146891} &  & \textcolor{black}{\small{}-324.165476} &  & \textcolor{black}{\small{}-324.172517}\tabularnewline
\textcolor{black}{\small{}14 } &  & \textcolor{black}{\small{}C$_{2}$H$_{6}$ } &  & \textcolor{black}{\small{}-79.603872} &  & \textcolor{black}{\small{}-79.628190} &  & \textcolor{black}{\small{}-79.631938}\tabularnewline
\textcolor{black}{\small{}15 } &  & \textcolor{black}{\small{}CH$_{3}$NH$_{2}$ } &  & \textcolor{black}{\small{}-95.631675} &  & \textcolor{black}{\small{}-95.655525} &  & \textcolor{black}{\small{}-95.659988}\tabularnewline
\textcolor{black}{\small{}16 } &  & \textcolor{black}{\small{}N$_{2}$H$_{4}$ } &  & \textcolor{black}{\small{}-111.629997} &  & \textcolor{black}{\small{}-111.655618} &  & \textcolor{black}{\small{}-111.661187}\tabularnewline
\textcolor{black}{\small{}17 } &  & \textcolor{black}{\small{}HOCl } &  & \textcolor{black}{\small{}-535.330974} &  & \textcolor{black}{\small{}-535.356795} &  & \textcolor{black}{\small{}-535.363297}\tabularnewline
\textcolor{black}{\small{}18 } &  & \textcolor{black}{\small{}C$_{3}$H$_{4}$ } &  & \textcolor{black}{\small{}-116.358023} &  & \textcolor{black}{\small{}-116.383694} &  & \textcolor{black}{\small{}-116.392120}\tabularnewline
\textcolor{black}{\small{}19 } &  & \textcolor{black}{\small{}CH$_{3}$Cl } &  & \textcolor{black}{\small{}-499.486888} &  & \textcolor{black}{\small{}-499.518970} &  & \textcolor{black}{\small{}-499.522907}\tabularnewline
\textcolor{black}{\small{}20 } &  & \textcolor{black}{\small{}CH$_{3}$SH } &  & \textcolor{black}{\small{}-438.083227} &  & \textcolor{black}{\small{}-438.117335} &  & \textcolor{black}{\small{}-438.123037}\tabularnewline
\textcolor{black}{\small{}21 } &  & \textcolor{black}{\small{}C$_{2}$FH$_{3}$ } &  & \textcolor{black}{\small{}-177.487929} &  & \textcolor{black}{\small{}-177.523874} &  & \textcolor{black}{\small{}-177.532492}\tabularnewline
\textcolor{black}{\small{}22 } &  & \textcolor{black}{\small{}CH$_{3}$OCH$_{3}$ } &  & \textcolor{black}{\small{}-154.679334} &  & \textcolor{black}{\small{}-154.720399} &  & \textcolor{black}{\small{}-154.727219}\tabularnewline
\textcolor{black}{\small{}23 } &  & \textcolor{black}{\small{}C$_{3}$H$_{6}$ } &  & \textcolor{black}{\small{}-117.568157} &  & \textcolor{black}{\small{}-117.614170} &  & \textcolor{black}{\small{}-117.620957}\tabularnewline
\textcolor{black}{\small{}24 } &  & \textcolor{black}{\small{}C$_{2}$H$_{4}$O } &  & \textcolor{black}{\small{}-153.450942} &  & \textcolor{black}{\small{}-153.496429} &  & \textcolor{black}{\small{}-153.504674}\tabularnewline
\textcolor{black}{\small{}25 } &  & \textcolor{black}{\small{}HCF$_{3}$ } &  & \textcolor{black}{\small{}-337.770010} &  & \textcolor{black}{\small{}-337.814755} &  & \textcolor{black}{\small{}-337.824683}\tabularnewline
\textcolor{black}{\small{}26 } &  & \textcolor{black}{\small{}C$_{2}$H$_{5}$N } &  & \textcolor{black}{\small{}-133.585955} &  & \textcolor{black}{\small{}-133.635764} &  & \textcolor{black}{\small{}-133.643763}\tabularnewline
\textcolor{black}{\small{}27 } &  & \textcolor{black}{\small{}COF$_{2}$ } &  & \textcolor{black}{\small{}-312.560599} &  & \textcolor{black}{\small{}-312.607818} &  & \textcolor{black}{\small{}-312.620274}\tabularnewline
\textcolor{black}{\small{}28 } &  & \textcolor{black}{\small{}CO$_{2}$ } &  & \textcolor{black}{\small{}-188.249006} &  & \textcolor{black}{\small{}-188.301643} &  & \textcolor{black}{\small{}-188.311990}\tabularnewline
\textcolor{black}{\small{}29 } &  & \textcolor{black}{\small{}OCS } &  & \textcolor{black}{\small{}-510.822787} &  & \textcolor{black}{\small{}-510.878940} &  & \textcolor{black}{\small{}-510.891218}\tabularnewline
\textcolor{black}{\small{}30 } &  & \textcolor{black}{\small{}BCl$_{3}$ } &  & \textcolor{black}{\small{}-1403.962819} &  & \textcolor{black}{\small{}-1404.036515} &  & \textcolor{black}{\small{}-1404.045130}\tabularnewline
\hline 
\end{tabular}
\end{table*}

\textcolor{black}{The situation is quite different in polyatomic systems
where the orbital localizability changes drastically the results obtained
with NOF-cMP2 and NOF-oiMP2. Both methods have been tested on a set
of 30 selected molecules with a dominant dynamic electron correlation
to compare with reliable MP2 energies. We must be aware that the applicability
of standard MP2 is restricted to cases without static correlation,
otherwise, we obtain an excess of correlation energy. An example is
the case of ozone which has an important multiconfigurational character.
In this case, NOF-oiMP2 predicts a total energy that is about 57 kcal/mol
higher than the MP2 value for the cc-pVTZ basis set \citep{Dunning1989}.
Consequently, an upper bound to the total MP2 energy can be expected
in most cases, since a fraction, however small, of non-dynamic correlation
is present.}

\textcolor{black}{The collection of total energies for the selected
set of molecules, calculated at their experimental geometries \citep{CCCBDB}
using the cc-pVTZ basis set \citep{Dunning1989}, can be found in
Table \ref{tab:Comparison}. For the whole set, the average differences
in the NOF-cMP2 and NOF-oiMP2 energies from MP2 are 34.5, and 6.3
mHartree, respectively. The data reveals an outstanding improvement
in the total energies of the NOF-oiMP2 over the NOF-cMP2.}

To summarize, it has been shown that a reformulation of the dynamic
electron-correlation energy based on the orbital-invariant MP2 allows
to extend the NOF-MP2 method to any type of orbitals, including the
typical localized orbitals of electron-pair-based NOFs. The global
character of the method was demonstrated in terms of relative and
total energies, since the dynamic and static correlation can be recovered
in one shot for any type of system, including weakly bound van der
Waals species.

\selectlanguage{american}%
\textbf{Acknowledgments:} The author thanks Fred Manby for his useful
suggestions during the ESCMQC2017. The SGI/IZO-SGIker of UPV/EHU supported
by European funding (ERDF and ESF) is gratefully acknowledged for
generous allocation of computational resources. The support from MINECO
(Grant No. CTQ2015- 67608-P) is also acknowledged.

\selectlanguage{english}%


\begin{thebibliography}{39}
\expandafter\ifx\csname natexlab\endcsname\relax\def\natexlab#1{#1}\fi
\expandafter\ifx\csname bibnamefont\endcsname\relax
  \def\bibnamefont#1{#1}\fi
\expandafter\ifx\csname bibfnamefont\endcsname\relax
  \def\bibfnamefont#1{#1}\fi
\expandafter\ifx\csname citenamefont\endcsname\relax
  \def\citenamefont#1{#1}\fi
\expandafter\ifx\csname url\endcsname\relax
  \def\url#1{\texttt{#1}}\fi
\expandafter\ifx\csname urlprefix\endcsname\relax\def\urlprefix{URL }\fi
\providecommand{\bibinfo}[2]{#2}
\providecommand{\eprint}[2][]{\url{#2}}

\bibitem[{\citenamefont{Valderrama et~al.}(1997)\citenamefont{Valderrama,
  Lude{\~{n}}a, and Hinze}}]{Valderrama1997}
\bibinfo{author}{\bibfnamefont{E.}~\bibnamefont{Valderrama}},
  \bibinfo{author}{\bibfnamefont{E.~V.} \bibnamefont{Lude{\~{n}}a}},
  \bibnamefont{and} \bibinfo{author}{\bibfnamefont{J.}~\bibnamefont{Hinze}},
  \bibinfo{journal}{J. Chem. Phys.} \textbf{\bibinfo{volume}{106}},
  \bibinfo{pages}{9227} (\bibinfo{year}{1997}).

\bibitem[{\citenamefont{Valderrama et~al.}(1999)\citenamefont{Valderrama,
  Lude{\~{n}}a, and Hinze}}]{Valderrama1999}
\bibinfo{author}{\bibfnamefont{E.}~\bibnamefont{Valderrama}},
  \bibinfo{author}{\bibfnamefont{E.~V.} \bibnamefont{Lude{\~{n}}a}},
  \bibnamefont{and} \bibinfo{author}{\bibfnamefont{J.}~\bibnamefont{Hinze}},
  \bibinfo{journal}{J. Chem. Phys.} \textbf{\bibinfo{volume}{110}},
  \bibinfo{pages}{2343} (\bibinfo{year}{1999}).

\bibitem[{\citenamefont{Piris}(2017)}]{Piris2017}
\bibinfo{author}{\bibfnamefont{M.}~\bibnamefont{Piris}},
  \bibinfo{journal}{Phys. Rev. Lett.} \textbf{\bibinfo{volume}{119}},
  \bibinfo{pages}{063002} (\bibinfo{year}{2017}).

\bibitem[{\citenamefont{Piris}(2018)}]{Piris2018}
\bibinfo{author}{\bibfnamefont{M.}~\bibnamefont{Piris}}, in
  \emph{\bibinfo{booktitle}{Many-body approaches at different scales: a tribute
  to N. H. March on the occasion of his 90th birthday}}, edited by
  \bibinfo{editor}{\bibfnamefont{G.~G.~N.} \bibnamefont{Angilella}}
  \bibnamefont{and} \bibinfo{editor}{\bibfnamefont{C.}~\bibnamefont{Amovilli}}
  (\bibinfo{publisher}{Springer}, \bibinfo{address}{New York},
  \bibinfo{year}{2018}), chap.~\bibinfo{chapter}{22}, pp.
  \bibinfo{pages}{283--300}.

\bibitem[{\citenamefont{Piris et~al.}(2013{\natexlab{a}})\citenamefont{Piris,
  Matxain, Lopez, and Ugalde}}]{Piris2013}
\bibinfo{author}{\bibfnamefont{M.}~\bibnamefont{Piris}},
  \bibinfo{author}{\bibfnamefont{J.~M.} \bibnamefont{Matxain}},
  \bibinfo{author}{\bibfnamefont{X.}~\bibnamefont{Lopez}}, \bibnamefont{and}
  \bibinfo{author}{\bibfnamefont{J.~M.} \bibnamefont{Ugalde}},
  \bibinfo{journal}{Theor. Chem. Acc.} \textbf{\bibinfo{volume}{132}},
  \bibinfo{pages}{1298} (\bibinfo{year}{2013}{\natexlab{a}}).

\bibitem[{\citenamefont{Matxain et~al.}(2012)\citenamefont{Matxain, Piris,
  Uranga, Lopez, Merino, and Ugalde}}]{Matxain2012a}
\bibinfo{author}{\bibfnamefont{J.~M.} \bibnamefont{Matxain}},
  \bibinfo{author}{\bibfnamefont{M.}~\bibnamefont{Piris}},
  \bibinfo{author}{\bibfnamefont{J.}~\bibnamefont{Uranga}},
  \bibinfo{author}{\bibfnamefont{X.}~\bibnamefont{Lopez}},
  \bibinfo{author}{\bibfnamefont{G.}~\bibnamefont{Merino}}, \bibnamefont{and}
  \bibinfo{author}{\bibfnamefont{J.~M.} \bibnamefont{Ugalde}},
  \bibinfo{journal}{ChemPhysChem} \textbf{\bibinfo{volume}{13}},
  \bibinfo{pages}{2297} (\bibinfo{year}{2012}).

\bibitem[{\citenamefont{Saebo}(2002)}]{Saebo2002}
\bibinfo{author}{\bibfnamefont{S.}~\bibnamefont{Saebo}}, in
  \emph{\bibinfo{booktitle}{Computational Chemistry: Reviews of Current Trends,
  Vol. 7}} (\bibinfo{year}{2002}), pp. \bibinfo{pages}{63--87}.

\bibitem[{\citenamefont{Werner et~al.}(2015)\citenamefont{Werner, Knizia,
  Krause, Schwilk, and Dornbach}}]{Werner2015}
\bibinfo{author}{\bibfnamefont{H.~J.} \bibnamefont{Werner}},
  \bibinfo{author}{\bibfnamefont{G.}~\bibnamefont{Knizia}},
  \bibinfo{author}{\bibfnamefont{C.}~\bibnamefont{Krause}},
  \bibinfo{author}{\bibfnamefont{M.}~\bibnamefont{Schwilk}}, \bibnamefont{and}
  \bibinfo{author}{\bibfnamefont{M.}~\bibnamefont{Dornbach}},
  \bibinfo{journal}{Journal of Chemical Theory and Computation}
  \textbf{\bibinfo{volume}{11}}, \bibinfo{pages}{484} (\bibinfo{year}{2015}).

\bibitem[{\citenamefont{Pulay}(1983)}]{Pulay1983}
\bibinfo{author}{\bibfnamefont{P.}~\bibnamefont{Pulay}},
  \bibinfo{journal}{Chem. Phys. Lett.} \textbf{\bibinfo{volume}{100}},
  \bibinfo{pages}{151} (\bibinfo{year}{1983}).

\bibitem[{\citenamefont{Zalesny et~al.}(2011)\citenamefont{Zalesny,
  Papadopoulos, Mezey, and Leszczynski}}]{LSTCQ2011}
\bibinfo{editor}{\bibfnamefont{R.}~\bibnamefont{Zalesny}},
  \bibinfo{editor}{\bibfnamefont{M.~G.} \bibnamefont{Papadopoulos}},
  \bibinfo{editor}{\bibfnamefont{P.~G.} \bibnamefont{Mezey}}, \bibnamefont{and}
  \bibinfo{editor}{\bibfnamefont{J.}~\bibnamefont{Leszczynski}}, eds.,
  \emph{\bibinfo{title}{Linear-Scaling Techniques in Computational Chemistry
  and Physics}} (\bibinfo{publisher}{Springer}, \bibinfo{address}{New York},
  \bibinfo{year}{2011}).

\bibitem[{\citenamefont{Piris}(2007)}]{Piris2007b}
\bibinfo{author}{\bibfnamefont{M.}~\bibnamefont{Piris}}, \bibinfo{journal}{Adv.
  Chem. Phys.} \textbf{\bibinfo{volume}{134}}, \bibinfo{pages}{387}
  (\bibinfo{year}{2007}).

\bibitem[{\citenamefont{Piris and Ugalde}(2014)}]{Piris2014a}
\bibinfo{author}{\bibfnamefont{M.}~\bibnamefont{Piris}} \bibnamefont{and}
  \bibinfo{author}{\bibfnamefont{J.~M.} \bibnamefont{Ugalde}},
  \bibinfo{journal}{Int. J. Quantum Chem.} \textbf{\bibinfo{volume}{114}},
  \bibinfo{pages}{1169} (\bibinfo{year}{2014}).

\bibitem[{\citenamefont{Coleman}(1963)}]{Coleman1963}
\bibinfo{author}{\bibfnamefont{A.~J.} \bibnamefont{Coleman}},
  \bibinfo{journal}{Rev. Mod. Phys.} \textbf{\bibinfo{volume}{35}},
  \bibinfo{pages}{668} (\bibinfo{year}{1963}).

\bibitem[{\citenamefont{Klyachko}(2006)}]{Klyachko2006}
\bibinfo{author}{\bibfnamefont{A.~A.} \bibnamefont{Klyachko}},
  \bibinfo{journal}{J. Phys.: Conf. Ser.} \textbf{\bibinfo{volume}{36}},
  \bibinfo{pages}{72} (\bibinfo{year}{2006}).

\bibitem[{\citenamefont{Altunbulak and Klyachko}(2008)}]{Altunbulak2008}
\bibinfo{author}{\bibfnamefont{M.}~\bibnamefont{Altunbulak}} \bibnamefont{and}
  \bibinfo{author}{\bibfnamefont{A.~A.} \bibnamefont{Klyachko}},
  \bibinfo{journal}{Comm. Math. Phys.} \textbf{\bibinfo{volume}{282}},
  \bibinfo{pages}{287} (\bibinfo{year}{2008}).

\bibitem[{\citenamefont{Valone}(1980)}]{Valone1980}
\bibinfo{author}{\bibfnamefont{S.~M.} \bibnamefont{Valone}},
  \bibinfo{journal}{J. Chem. Phys.} \textbf{\bibinfo{volume}{73}},
  \bibinfo{pages}{1344} (\bibinfo{year}{1980}).

\bibitem[{\citenamefont{Nguyen-Dang et~al.}(1985)\citenamefont{Nguyen-Dang,
  Lude{\~{n}}a, and Tal}}]{Nguyen-Dang1985}
\bibinfo{author}{\bibfnamefont{T.~T.} \bibnamefont{Nguyen-Dang}},
  \bibinfo{author}{\bibfnamefont{E.~V.} \bibnamefont{Lude{\~{n}}a}},
  \bibnamefont{and} \bibinfo{author}{\bibfnamefont{Y.}~\bibnamefont{Tal}},
  \bibinfo{journal}{J. Mol. Struc.: THEOCHEM} \textbf{\bibinfo{volume}{120}},
  \bibinfo{pages}{247} (\bibinfo{year}{1985}).

\bibitem[{\citenamefont{Mazziotti}(2012)}]{Mazziotti2012}
\bibinfo{author}{\bibfnamefont{D.~A.} \bibnamefont{Mazziotti}},
  \bibinfo{journal}{Phys. Rev. Lett.} \textbf{\bibinfo{volume}{108}},
  \bibinfo{pages}{263002} (\bibinfo{year}{2012}).

\bibitem[{\citenamefont{Piris}(2006)}]{Piris2006}
\bibinfo{author}{\bibfnamefont{M.}~\bibnamefont{Piris}}, \bibinfo{journal}{Int.
  J. Quantum Chem.} \textbf{\bibinfo{volume}{106}}, \bibinfo{pages}{1093}
  (\bibinfo{year}{2006}).

\bibitem[{\citenamefont{Mazziotti}(1998)}]{Mazziotti1998}
\bibinfo{author}{\bibfnamefont{D.~A.} \bibnamefont{Mazziotti}},
  \bibinfo{journal}{Chem. Phys. Lett.} \textbf{\bibinfo{volume}{289}},
  \bibinfo{pages}{419} (\bibinfo{year}{1998}).

\bibitem[{\citenamefont{Piris}(1999)}]{Piris1999}
\bibinfo{author}{\bibfnamefont{M.}~\bibnamefont{Piris}}, \bibinfo{journal}{J.
  Math. Chem.} \textbf{\bibinfo{volume}{25}}, \bibinfo{pages}{47}
  (\bibinfo{year}{1999}).

\bibitem[{\citenamefont{Piris}(2013{\natexlab{a}})}]{Piris2013b}
\bibinfo{author}{\bibfnamefont{M.}~\bibnamefont{Piris}}, \bibinfo{journal}{Int.
  J. Quantum Chem.} \textbf{\bibinfo{volume}{113}}, \bibinfo{pages}{620}
  (\bibinfo{year}{2013}{\natexlab{a}}).

\bibitem[{\citenamefont{Piris}(2014)}]{Piris2014c}
\bibinfo{author}{\bibfnamefont{M.}~\bibnamefont{Piris}}, \bibinfo{journal}{J.
  Chem. Phys.} \textbf{\bibinfo{volume}{141}}, \bibinfo{pages}{044107}
  (\bibinfo{year}{2014}).

\bibitem[{\citenamefont{Piris et~al.}(2011)\citenamefont{Piris, Lopez,
  Ruip\'{e}rez, Matxain, and Ugalde}}]{Piris2011}
\bibinfo{author}{\bibfnamefont{M.}~\bibnamefont{Piris}},
  \bibinfo{author}{\bibfnamefont{X.}~\bibnamefont{Lopez}},
  \bibinfo{author}{\bibfnamefont{F.}~\bibnamefont{Ruip\'{e}rez}},
  \bibinfo{author}{\bibfnamefont{J.~M.} \bibnamefont{Matxain}},
  \bibnamefont{and} \bibinfo{author}{\bibfnamefont{J.~M.}
  \bibnamefont{Ugalde}}, \bibinfo{journal}{J. Chem. Phys.}
  \textbf{\bibinfo{volume}{134}}, \bibinfo{pages}{164102}
  (\bibinfo{year}{2011}).

\bibitem[{\citenamefont{Piris et~al.}(2013{\natexlab{b}})\citenamefont{Piris,
  Matxain, and Lopez}}]{Piris2013e}
\bibinfo{author}{\bibfnamefont{M.}~\bibnamefont{Piris}},
  \bibinfo{author}{\bibfnamefont{J.~M.} \bibnamefont{Matxain}},
  \bibnamefont{and} \bibinfo{author}{\bibfnamefont{X.}~\bibnamefont{Lopez}},
  \bibinfo{journal}{J. Chem. Phys.} \textbf{\bibinfo{volume}{139}},
  \bibinfo{pages}{234109} (\bibinfo{year}{2013}{\natexlab{b}}).

\bibitem[{\citenamefont{Pernal}(2013)}]{Pernal2013}
\bibinfo{author}{\bibfnamefont{K.}~\bibnamefont{Pernal}},
  \bibinfo{journal}{Comp. Theor. Chem.} \textbf{\bibinfo{volume}{1003}},
  \bibinfo{pages}{127} (\bibinfo{year}{2013}).

\bibitem[{\citenamefont{Piris}(2013{\natexlab{b}})}]{Piris2013c}
\bibinfo{author}{\bibfnamefont{M.}~\bibnamefont{Piris}}, \bibinfo{journal}{J.
  Chem. Phys.} \textbf{\bibinfo{volume}{139}}, \bibinfo{pages}{064111}
  (\bibinfo{year}{2013}{\natexlab{b}}).

\bibitem[{\citenamefont{Piris et~al.}(2009)\citenamefont{Piris, Matxain, Lopez,
  and Ugalde}}]{Piris2009}
\bibinfo{author}{\bibfnamefont{M.}~\bibnamefont{Piris}},
  \bibinfo{author}{\bibfnamefont{J.~M.} \bibnamefont{Matxain}},
  \bibinfo{author}{\bibfnamefont{X.}~\bibnamefont{Lopez}}, \bibnamefont{and}
  \bibinfo{author}{\bibfnamefont{J.~M.} \bibnamefont{Ugalde}},
  \bibinfo{journal}{J. Chem. Phys.} \textbf{\bibinfo{volume}{131}},
  \bibinfo{pages}{021102} (\bibinfo{year}{2009}).

\bibitem[{\citenamefont{Piris et~al.}(2010)\citenamefont{Piris, Matxain, Lopez,
  and Ugalde}}]{Piris2010a}
\bibinfo{author}{\bibfnamefont{M.}~\bibnamefont{Piris}},
  \bibinfo{author}{\bibfnamefont{J.~M.} \bibnamefont{Matxain}},
  \bibinfo{author}{\bibfnamefont{X.}~\bibnamefont{Lopez}}, \bibnamefont{and}
  \bibinfo{author}{\bibfnamefont{J.~M.} \bibnamefont{Ugalde}},
  \bibinfo{journal}{J. Chem. Phys.} \textbf{\bibinfo{volume}{133}},
  \bibinfo{pages}{111101} (\bibinfo{year}{2010}).

\bibitem[{\citenamefont{Lowdin and Shull}(1955)}]{Lowdin1955d}
\bibinfo{author}{\bibfnamefont{P.~O.} \bibnamefont{Lowdin}} \bibnamefont{and}
  \bibinfo{author}{\bibfnamefont{H.}~\bibnamefont{Shull}},
  \bibinfo{journal}{Phys. Rev.} \textbf{\bibinfo{volume}{101}},
  \bibinfo{pages}{1730} (\bibinfo{year}{1955}).

\bibitem[{\citenamefont{Mitxelena et~al.}(2018)\citenamefont{Mitxelena,
  Rodr\'{i}guez-Mayorga, and Piris}}]{mitxelena2018a}
\bibinfo{author}{\bibfnamefont{I.}~\bibnamefont{Mitxelena}},
  \bibinfo{author}{\bibfnamefont{M.}~\bibnamefont{Rodr\'{i}guez-Mayorga}},
  \bibnamefont{and} \bibinfo{author}{\bibfnamefont{M.}~\bibnamefont{Piris}},
  \bibinfo{journal}{European Physics Journal B} \textbf{\bibinfo{volume}{91}},
  \bibinfo{pages}{109} (\bibinfo{year}{2018}).

\bibitem[{\citenamefont{{Johnson III}}(2016)}]{nist}
\bibinfo{editor}{\bibfnamefont{R.~D.} \bibnamefont{{Johnson III}}}, ed.,
  \emph{\bibinfo{title}{{NIST Standard Reference Database Number 101, Release
  18}}} (\bibinfo{year}{2016}).

\bibitem[{\citenamefont{{Chase, Jr.}}(1998)}]{Chase1998}
\bibinfo{author}{\bibfnamefont{M.~W.} \bibnamefont{{Chase, Jr.}}},
  \bibinfo{journal}{J. Phys. Chem. Ref. Data Monogr.}
  \textbf{\bibinfo{volume}{9}}, \bibinfo{pages}{1} (\bibinfo{year}{1998}).

\bibitem[{\citenamefont{Dunning~Jr.}(1989)}]{Dunning1989}
\bibinfo{author}{\bibfnamefont{T.~H.} \bibnamefont{Dunning~Jr.}},
  \bibinfo{journal}{J. Chem. Phys.} \textbf{\bibinfo{volume}{90}},
  \bibinfo{pages}{1007} (\bibinfo{year}{1989}).

\bibitem[{\citenamefont{Pulay and Saebo}(1986)}]{Pulay1986}
\bibinfo{author}{\bibfnamefont{P.}~\bibnamefont{Pulay}} \bibnamefont{and}
  \bibinfo{author}{\bibfnamefont{S.}~\bibnamefont{Saebo}},
  \bibinfo{journal}{Theor. Chim. Acta} \textbf{\bibinfo{volume}{69}},
  \bibinfo{pages}{357} (\bibinfo{year}{1986}).

\bibitem[{\citenamefont{Ogilvie and Wang}(1992)}]{Ogilvie1992}
\bibinfo{author}{\bibfnamefont{J.}~\bibnamefont{Ogilvie}} \bibnamefont{and}
  \bibinfo{author}{\bibfnamefont{F.~Y.} \bibnamefont{Wang}},
  \bibinfo{journal}{J. Mol. Struc.} \textbf{\bibinfo{volume}{273}},
  \bibinfo{pages}{277} (\bibinfo{year}{1992}).

\bibitem[{\citenamefont{Kendall et~al.}(1992)\citenamefont{Kendall, {Dunning,
  Jr.}, and Harrison}}]{Kendall1992}
\bibinfo{author}{\bibfnamefont{R.~A.} \bibnamefont{Kendall}},
  \bibinfo{author}{\bibfnamefont{T.~H.} \bibnamefont{{Dunning, Jr.}}},
  \bibnamefont{and} \bibinfo{author}{\bibfnamefont{R.~J.}
  \bibnamefont{Harrison}}, \bibinfo{journal}{J. Chem. Phys.}
  \textbf{\bibinfo{volume}{96}}, \bibinfo{pages}{6796} (\bibinfo{year}{1992}).

\bibitem[{\citenamefont{Woon and {Dunning, Jr.}}(1994)}]{Woon1994}
\bibinfo{author}{\bibfnamefont{D.~E.} \bibnamefont{Woon}} \bibnamefont{and}
  \bibinfo{author}{\bibfnamefont{T.~H.} \bibnamefont{{Dunning, Jr.}}},
  \bibinfo{journal}{J. Chem. Phys.} \textbf{\bibinfo{volume}{100}},
  \bibinfo{pages}{2975} (\bibinfo{year}{1994}).

\bibitem[{CCC()}]{CCCBDB}
\bibinfo{note}{{NIST Computational Chemistry Comparison and Benchmark
  Database}, NIST Standard Reference Database Number 101, Release 18, October
  2016, Editor: Russell D. Johnson III. Available at http://cccbdb.nist.gov/}.

\end{thebibliography}
\end{document}